\PassOptionsToPackage{unicode}{hyperref}
\PassOptionsToPackage{hyphens}{url}
\PassOptionsToPackage{dvipsnames,svgnames,x11names}{xcolor}
\documentclass[
  a4paper,
]{article}
\usepackage{xcolor}
\usepackage{amsmath,amssymb}
\usepackage{iftex}
\ifPDFTeX
  \usepackage[T1]{fontenc}
  \usepackage[utf8]{inputenc}
  \usepackage{textcomp} 
\else 
  \usepackage{unicode-math} 
  \defaultfontfeatures{Scale=MatchLowercase}
  \defaultfontfeatures[\rmfamily]{Ligatures=TeX,Scale=1}
\fi
\usepackage{lmodern}
\ifPDFTeX\else
\fi
\IfFileExists{upquote.sty}{\usepackage{upquote}}{}
\IfFileExists{microtype.sty}{
  \usepackage[]{microtype}
  \UseMicrotypeSet[protrusion]{basicmath} 
}{}
\usepackage{setspace}
\makeatletter
\@ifundefined{KOMAClassName}{
  \IfFileExists{parskip.sty}{%
    \usepackage{parskip}
  }{
    \setlength{\parindent}{0pt}
    \setlength{\parskip}{6pt plus 2pt minus 1pt}}
}{
  \KOMAoptions{parskip=half}}
\makeatother
\makeatletter
\ifx\paragraph\undefined\else
  \let\oldparagraph\paragraph
  \renewcommand{\paragraph}{
    \@ifstar
      \xxxParagraphStar
      \xxxParagraphNoStar
  }
  \newcommand{\xxxParagraphStar}[1]{\oldparagraph*{#1}\mbox{}}
  \newcommand{\xxxParagraphNoStar}[1]{\oldparagraph{#1}\mbox{}}
\fi
\ifx\subparagraph\undefined\else
  \let\oldsubparagraph\subparagraph
  \renewcommand{\subparagraph}{
    \@ifstar
      \xxxSubParagraphStar
      \xxxSubParagraphNoStar
  }
  \newcommand{\xxxSubParagraphStar}[1]{\oldsubparagraph*{#1}\mbox{}}
  \newcommand{\xxxSubParagraphNoStar}[1]{\oldsubparagraph{#1}\mbox{}}
\fi
\makeatother

\usepackage{longtable,booktabs,array}
\usepackage{calc} 
\usepackage{etoolbox}
\makeatletter
\patchcmd\longtable{\par}{\if@noskipsec\mbox{}\fi\par}{}{}
\makeatother
\IfFileExists{footnotehyper.sty}{\usepackage{footnotehyper}}{\usepackage{footnote}}
\makesavenoteenv{longtable}
\usepackage{graphicx}
\makeatletter
\newsavebox\pandoc@box
\newcommand*\pandocbounded[1]{
  \sbox\pandoc@box{#1}%
  \Gscale@div\@tempa{\textheight}{\dimexpr\ht\pandoc@box+\dp\pandoc@box\relax}%
  \Gscale@div\@tempb{\linewidth}{\wd\pandoc@box}%
  \ifdim\@tempb\p@<\@tempa\p@\let\@tempa\@tempb\fi
  \ifdim\@tempa\p@<\p@\scalebox{\@tempa}{\usebox\pandoc@box}%
  \else\usebox{\pandoc@box}%
  \fi%
}
\def\fps@figure{htbp}
\makeatother

\NewDocumentCommand\citeproctext{}{}

\makeatletter
 \let\@cite@ofmt\@firstofone
 \def\@biblabel#1{}
 \def\@cite#1#2{{#1\if@tempswa , #2\fi}}
\makeatother
\newlength{\cslhangindent}
\newlength{\csllabelwidth}
\newenvironment{CSLReferences}[2] 
 {\begin{list}{}{%
  \setlength{\itemindent}{0pt}
  \setlength{\leftmargin}{0pt}
  \setlength{\parsep}{0pt}
  \ifodd #1
   \setlength{\leftmargin}{\cslhangindent}
   \setlength{\itemindent}{-1\cslhangindent}
  \fi
  \setlength{\itemsep}{#2\baselineskip}}}
 {\end{list}}
\usepackage{calc}

\usepackage[margin=1in]{geometry}
\usepackage{setspace}
\usepackage{xcolor}
\usepackage{graphicx}
\usepackage{float}
\usepackage{tcolorbox}
\usepackage{wrapfig}
\usepackage{booktabs}
\usepackage{multirow}
\usepackage{multicol}
\usepackage{makecell}
\usepackage{siunitx}
\usepackage{tabularx}

\usepackage{amsmath, amssymb, amsfonts}
\usepackage{hyperref}
\usepackage[backend=biber]{biblatex}
\usepackage{lineno}

\usepackage{fancyhdr}
\fancypagestyle{plain}{
  \fancyhf{}
  \fancyhead[L,R]{\small\color{red}PREPRINT/WORKING PAPER}
  \fancyhead[C]{%
    \small\color{red}\url{https://krishpn.github.io}
  }
  \fancyfoot[L,R]{\small\color{red}PREPRINT/WORKING PAPER}
  \fancyfoot[C]{\thepage}
}

\makeatletter
\@ifpackageloaded{caption}{}{\usepackage{caption}}
\AtBeginDocument{%
\ifdefined\contentsname
  \renewcommand*\contentsname{Table of contents}
\else
  \newcommand\contentsname{Table of contents}
\fi
\ifdefined\listfigurename
  \renewcommand*\listfigurename{List of Figures}
\else
  \newcommand\listfigurename{List of Figures}
\fi
\ifdefined\listtablename
  \renewcommand*\listtablename{List of Tables}
\else
  \newcommand\listtablename{List of Tables}
\fi
\ifdefined\figurename
  \renewcommand*\figurename{Figure}
\else
  \newcommand\figurename{Figure}
\fi
\ifdefined\tablename
  \renewcommand*\tablename{Table}
\else
  \newcommand\tablename{Table}
\fi
}
\@ifpackageloaded{float}{}{\usepackage{float}}
\floatstyle{ruled}
\@ifundefined{c@chapter}{\newfloat{codelisting}{h}{lop}}{\newfloat{codelisting}{h}{lop}[chapter]}
\floatname{codelisting}{Listing}

\makeatother
\makeatletter
\@ifpackageloaded{caption}{}{\usepackage{caption}}
\@ifpackageloaded{subcaption}{}{\usepackage{subcaption}}
\makeatother
\usepackage{bookmark}
\IfFileExists{xurl.sty}{\usepackage{xurl}}{} 
\hypersetup{
  pdftitle={The Strategic Gap: How AI-Driven Timing and Complexity Shape Investor Trust in the Age of Digital Agents},
  pdfauthor={Krishna Neupane     ORCID: 0000-0003-1689-0557},
  colorlinks=true,
  linkcolor={blue},
  filecolor={Maroon},
  citecolor={Blue},
  urlcolor={Blue},
  pdfcreator={LaTeX via pandoc}}

\title{The Strategic Gap: How AI-Driven Timing and Complexity Shape
Investor Trust in the Age of Digital Agents}
\author{Krishna Neupane \newline \vspace{0.5em}
\small \href{mailto:kneupan@gmu.edu}{kneupan@gmu.edu} \vspace{0.5em}
\small ORCID: 0000-0003-1689-0557}
\date{2026-02-19}
\begin{document}
\maketitle
\begin{abstract}
\noindent Traditional models of market efficiency assume that equity
prices incorporate information based on content alone, often neglecting
the structural influence of reporting timing and cadence. This study
introduces the Autonomous Disclosure Regulator, a multi-node AI
framework designed to audit the intersection of disclosure complexity
and filing unpredictability. Analyzing a population of 484,796
regulatory filings, the research identifies a structural Strategic Gap:
a state where companies use confusing language and unpredictable timing
to slow down how fast the market learns the truth by 60\%. The results
demonstrate a fundamental computational asymmetry in contemporary
capital markets. While investors are now good at spotting ``copy-paste''
text, they remain vulnerable to strategic timing that obscures
structural deterioration. The framework isolates 39 high-priority
failures where the convergence of dense text and temporal surprises
facilitated significant information rent extraction by insiders. By
implementing a recursive agentic audit, the system identifies a
cumulative welfare recovery potential of over 360\% and demonstrates
near-perfect resilience against technical data interruptions. The study
concludes by proposing a transition toward an agentic regulatory state,
arguing that as information integration costss rise, infrastructure must
evolve from passive data repositories into active auditing nodes capable
of real-time synthesis to preserve market integrity.

\vspace{1em}

\noindent © Krishna Neupane 2026. All rights reserved. This working
paper is part of on going review for publication. No part of this
publication may be reproduced without prior permission.
\url{https://krishpn.github.io}

\vspace{1em}

\noindent \textbf{Keywords:} Agentic AI, Computational Asymmetry,
RegTech, Market Microstructure, Large Language Models in Finance,
Strategic Disclosure, SEC, Form 10-K, Form 10-Q

\noindent \textbf{JEL Classification:} G14, K22, C45, D82
\end{abstract}

\setstretch{1.5}
\section{Introduction}\label{sec:introduction}

The First Theorem of Welfare Economics assumes a frictionless
information environment where competitive equilibria are Pareto optimal.
However, this study argues that the theorem is structurally challenged
by asymmetric processing capacities. While traditional theory assumes
information is fully incorporated into prices, the theory of Rational
Inattention proposed by Sims (2003) models how agents strategically
simplify information under cognitive constraints, creating frictions
that prevent full price integration of Hirshleifer and Luo (2001). The
fundamental paradox established by Grossman and Stiglitz (1980)---that
markets cannot achieve perfect efficiency because information collection
is costly---is intensified by high-velocity, automated reporting. As
noted by Subramanyam (1996), managers utilize the discretionary nature
of reporting to produce standardized text with low informational density
that leverages the limited processing bandwidth of market participants.

The primary objective of this research is to quantify the extent to
which modern disclosure patterns impede price discovery. To achieve
this, the study develops the
\textbf{Autonomous Disclosure Regulator (ADR)}, a system designed to
simulate and audit the informedness gap between automated filing
production and human regulatory oversight. By analyzing 484,796 SEC
filings, the study seeks to determine if a state of symmetric
computational equilibrium can be restored through autonomous auditing.

The analysis isolates a structural failure in price discovery termed the
\textbf{``Strategic Gap.''} In this regime---defined by high semantic
friction (\(Z_{Comp}\)) and low predictability (\(\Phi\))---the study
observes a 60\% reduction in the Price Discovery Rate (\(V\)), with
values declining from 3.07 to 1.23. This latency allows for regulatory
arbitrage that renders traditional oversight windows, such as the
Section 16 ``Two-Day Rule,'\,' functionally ineffective. Furthermore,
the research identifies 39 cases of \textbf{Non-Informative Compliance}.
These represent filings that maintain legal adherence to guidelines but
are devoid of the idiosyncratic data necessary for fundamental
valuation. The findings suggest that this noise is strategically
deployed to facilitate insider rent-seeking, resulting in a calculated
Welfare Gain Potential (\(\Gamma\)) of 360,050\% through improved
oversight.

The study provides a technical contribution by operationalizing the
\textbf{Agentic Regulatory Auditor (ARA)}. This architecture shifts the
cost-curve of oversight from manual review to automated vigilance
through a four-node modular structure:

\begin{itemize}
    \item \textbf{Node A (Semantic Interface):} Achieves 98.11\% structural extraction accuracy, identifying data within standardized text using high-dimensional transformer embeddings.
    \item \textbf{Node B (Temporal Syntax):} Measures the entropy of disclosure cadence to identify the 23\% of filings within the Strategic Gap.
    \item \textbf{Node C (System Persistence):} Employs stateful audit trails to resolve data-stream interruptions, achieving 99.98\% system resilience.
    \item \textbf{Node D (Recursive Search):} Functions as a validation protocol using recursive synthesis to narrow the investigative scope with 20.88\% precision.
\end{itemize}

The paper concludes by proposing the
\textbf{Agent-First Disclosure Mandate (ADS)}. This policy moves toward
a regulatory state where filing validity is contingent upon immediate
machine-readability, thereby mitigating the computational divergence
between market participants.

\section{Related Literature}\label{sec:related_literature}

The framework developed in this study builds upon, yet fundamentally
departs from, the ``Disclosure Processing Cost'' paradigm synthesized by
Blankespoor, deHaan, and Marinovic (2020). In the traditional view,
disclosure frictions---comprising awareness, acquisition, and
integration costs---are treated as equilibrium-stabilizing mechanisms
that rationalize market phenomena such as post-earnings announcement
drift. Under this equilibrium, the ``Integration Step'' is viewed as a
cognitive burden borne by investors in exchange for competitive returns.
This research challenges that passive characterization, redefining these
costs as a structural regulatory failure that can be mitigated through
agentic intervention. The primary obstacle to such intervention,
however, has been the technological inability to automate the
``Integration Step'' without losing the nuanced context of financial
reporting.

\subsection{From Lexical Dictionaries to Semantic Embeddings}

To address this automation deficit, the literature first sought to
quantify textual information through dictionary-based methods. The
seminal work of Loughran and Wellman (2011) revolutionized financial
natural language processing by demonstrating that standard
psycholinguistic dictionaries were poorly suited for the nuanced jargon
of capital markets. While this lexicon-based approach reduced
acquisition costs, it struggled to scale alongside the subsequent rise
in Business Model Complexity identified by Loughran and McDonald (2024).
This literature argues that as disclosures become increasingly
voluminous and integrated, the mere presence of negative keywords is no
longer a sufficient signal for risk.

This study addresses the limitations of keyword frequency by
operationalizing transformer-based architectures capable of capturing
context-dependent meaning. The present framework addresses this
informational density by moving beyond frequency-based word-count
paradigms. Node A utilizes high-dimensional transformer embeddings to
detect Semantic Friction (\(Z_{Comp}\)), an approach consistent with the
``structural staleness'' logic of Cohen, Malloy, and Nguyen (2020). The
observed 0.082 Spearman correlation between complexity and sentiment
provides empirical evidence for the Strategic Obfuscation Hypothesis:
managers may utilize complex, neutral jargon to establish a baseline of
information noise. This baseline effectively masks material shocks
within routine boilerplate, rendering legacy dictionary-based audits
insufficient for detecting idiosyncratic risk. However, identifying
semantic friction is only the first stage of regulatory intervention;
the critical challenge lies in determining how these frictions are
utilized to mask systemic instability.

\subsection{Re-characterizing Integration Costs as an Agentic Pivot}

To transition from mere detection to active oversight, this study shifts
the analytical objective from price prediction to structural auditing.
This study further departs from the Blankespoor, deHaan, and Marinovic
(2020) paradigm by shifting the analytical objective of disclosure
processing. While traditional literature focuses on how processing costs
affect investor choice and price efficiency, the current framework
prioritizes the structural anatomy of the disclosure. Node D (Recursive
Synthesis) is designed not to predict price drift, but to autonomously
identify latent structural breaks---defined here as Nominal
Transparency---that obscure underlying insolvency or reporting
irregularities. By focusing on the structural properties of the signal
rather than the market's aggregate reaction, the regulatory focus shifts
from ex-post enforcement to ex-ante detection.

By focusing on the structural properties of the signal rather than the
market's aggregate reaction, the regulatory focus shifts from ex-post
enforcement to ex-ante detection. This shift introduces the Agentic
Pivot in Information Theory. Current literature largely envisions a
dyadic relationship between a firm (the signal producer) and a
capacity-constrained investor (the processor).This paper introduces a
third, dominant actor: the Autonomous Agent. In this triadic model, the
regulator is no longer a passive observer of market frictions but an
active participant using transformer-based encoders to mitigate
standardized disclosure opacity. This transition from human-led
monitoring to machine-led verification implies that informational
frictions are no longer inevitable market features, but addressable
system vulnerabilities.

\subsection{Information Asymmetry as a Strategic Choice}

By treating these frictions as addressable vulnerabilities, the study
must then address whether their existence is accidental or intentional.
Finally, this work challenges the assumption that integration costs are
a natural byproduct of business complexity. Following the logic of
Subramanyam (1996), the analysis demonstrates that strategic asymmetry
is frequently a deliberate managerial choice. By isolating the 39
``Critical Failures'' in the sample, the study provides evidence that
managers exploit the intersection of linguistic density and temporal
variance to create Strategic Gaps.

Unlike the participants in Tetlock (2008) who react to stale news,
actors in these identified gaps lack Temporal Syntax. This observation
supports the proposal for an Agent-First Disclosure Mandate (ADS). This
policy suggests that in an environment of high agentic velocity, the
validity of a public filing should be contingent upon its immediate
machine-ingestibility at \(T+0\), thereby neutralizing the rent-seeking
windows that exist behind high-friction, erratic disclosures.

\begin{table}[htbp]
\centering
\caption{Mapping Multi-Disciplinary Market Frictions to ARA Architectural Nodes}
\label{tab:theoretical_roadmap}
\resizebox{\textwidth}{!}{%
\begin{tabular}{|l|l|p{3.5cm}|p{3.5cm}|p{4.5cm}|}
\hline
\textbf{Paradigm} & \textbf{Leading Author(s)} & \textbf{Mechanism / Friction} & \textbf{Regulator's State} & \textbf{The ARA Solution (Node)} \\ \hline
\textbf{Welfare} & Grossman \& Stiglitz (1980) & \textbf{Information Paradox:} Costly signal acquisition. & Informational Asymmetry & \textbf{Node A/B:} Near-zero marginal cost extraction. \\ \hline
\textbf{Inattention} & Sims (2003) & \textbf{Rational Inattention:} Strategic simplification of data. & Processing Constraints & \textbf{Node B:} Filters temporal entropy via Chronos-Small auditing. \\ \hline
\textbf{Behavioral} & Hirshleifer (2001) & \textbf{Limited Attention:} Market neglect of non-salient data. & Processing Latency & \textbf{Node C:} LangGraph Persistence maintains audit continuity. \\ \hline
\textbf{Textual} & Loughran \& McDonald (2024) & \textbf{Business Complexity:} Strategic jargon as noise. & Interpretive Bias & \textbf{Node A:} Vector embeddings replace static dictionaries. \\ \hline
\textbf{Reporting} & Subramanyam (1996) & \textbf{Discretionary Smoothing:} Use of noise to mask quality. & Integration Friction & \textbf{Node A/D:} Complexity auditing identifies smoothing. \\ \hline
\textbf{Microstructure} & Kyle (1985) & \textbf{Strategic Timing:} Insiders hiding in automated noise. & Strategic Information Gap & \textbf{Node D/C:} Maps noise to insider trading windows. \\ \hline
\textbf{AI-Economics} & Blankespoor (2025) & \textbf{Automated Obfuscation:} Machine-generated tone spin. & Automated Processing Fog & \textbf{Node D:} Recursive Synthesis validates structural signals. \\ \hline
\textbf{This Paper} & \textbf{Neupane (2026)} & \textbf{Agentic Velocity} & \textbf{Agentic Vigilance} & \textbf{Full Framework:} Machine-speed Welfare restoration. \\ \hline
\end{tabular}%
}
\end{table}

As illustrated in Table \ref{tab:theoretical_roadmap}, the proposed
framework does not replace existing economic paradigms; rather, it
provides a specific computational node designed to resolve each
identified friction. By transitioning from a regime of processing
inertia toward a state of Agentic Vigilance, this research offers a
technical solution to modern information failures. The framework
demonstrates that by neutralizing computational divergence, it is
possible to restore the First Theorem of Welfare Economics, ensuring
that competitive equilibria remain Pareto optimal in an environment of
automated disclosure.

\section{Theoretical Framework}\label{sec:theoretical_framework}

A core tenet of this framework is that the Informedness Gap (\(\Psi\))
is a dynamic variable driven by the strategic reallocation of
informational capacity and the emergence of Agentic Velocity. The study
adapts the attention allocation model proposed by Mondria and
Quintana-Domeque (2013), which theorizes that financial contagion occurs
when investors relocate limited attention between sectors, thereby
increasing information asymmetry in neglected areas. This research
extends that logic to the regulatory domain: when oversight bodies
concentrate on high-volatility events, such as busy earnings days, they
effectively relocate attention away from firms characterized by high
Semantic Friction (\(Z_{Comp}\)) and low Agentic Predictability
(\(\Phi\)).

\subsection{The Three Pillars of Agentic Velocity}\label{sec:three_pillars_agentic_velocity}

The transition from human-mediated disclosures to high-frequency
reporting environments necessitates a new taxonomy of friction. This
study defines three critical dimensions---summarized in Table
\ref{tab:pillars_velocity}---that collectively drive the modern
``Strategic Gap.'' These pillars represent the structural mechanisms
used to exploit the attention constraints identified in the Mondria and
Quintana-Domeque (2013) model.

\begin{table}[htbp]
\centering
\caption{Conceptual Framework of Agentic Velocity: Mapping Dimensional Frictions to Regulatory Intervention Nodes}
\label{tab:pillars_velocity}
\renewcommand{\arraystretch}{1.5}
\begin{tabularx}{\textwidth}{l p{2.5cm} X c}
\toprule
\textbf{Dimension} & \textbf{Definition} & \textbf{Welfare Economics Implication} & \textbf{ARA Node} \\
\midrule
Filing Latency & Temporal delta between transaction and SEC receipt. & \textbf{Friction:} Creates front-running windows where information is technically public but practically inaccessible. & \textbf{Node B} \\ Processing Velocity & Time required for semantic parsing of raw \texttt{.txt} files. & \textbf{Informedness Gap:} Automated parsing at 22x speed renders manual analysis stale at $T+0$. & \textbf{Node A} \\
Decision Autonomy & Degree of execution without human-in-the-loop. & \textbf{Market Failure:} High-velocity autonomy facilitates synchronized herding and flash volatility at the $T+100ms$ scale. & \textbf{Node C/D} \\
\bottomrule
\end{tabularx}

\subcaption*{\textit{Note:} This table establishes the three-dimensional ``Agentic Velocity'' framework used to audit modern market failures. Each dimension identifies a specific friction where human-speed oversight is mathematically insufficient to maintain Pareto-optimal discovery. The ARA framework addresses these failures by deploying Node B (Temporal Syntax) to audit latency entropy, Node A (Semantic Interface) to neutralize the processing gap via high-dimensional embeddings, and Nodes C and D (Recursive Supervision) to provide stateful, autonomous oversight of high-autonomy reporting regimes.}
\end{table}

The first dimension, \textbf{Filing Latency}, represents the temporal
delta between a transaction and its official receipt by the SEC. In a
regime of Agentic Velocity, while the legal reporting window may remain
constant, the effective window for institutional insiders is compressed.
This creates a ``front-running'' opportunity where information is
technically public but practically inaccessible to manual processors.

The second dimension, \textbf{Processing Velocity}, quantifies the speed
at which a semantic processor can transform raw text into a structured
trade signal. The study observes that automated systems parse
disclosures at roughly 22x the speed of human analysts. This speed
differential transforms the ``Integration Step'' from a shared market
cost into a structural Informedness Gap, rendering any manually
processed signal ``stale'' by the time it reaches the order book.

Finally, \textbf{Decision Autonomy} describes the degree to which these
automated systems execute outcomes without direct human intervention.
When high velocity is paired with high autonomy, the risk of
synchronized herding increases. At the \(T+100ms\) scale, the lack of a
``human-in-the-loop'' can trigger flash volatility, representing a
significant departure from Pareto-optimal discovery and a primary
failure in market microstructure.

\subsection{The Phase Shift toward Agentic Vigilance}\label{sec:the_phase_shift_toward_agentic_vigilance}

The structural velocity of modern reporting facilitates a novel form of
\textbf{Regulatory Arbitrage}. Analysis suggests that managers operating
within the Strategic Gap quadrant exploit the predictable relocation of
oversight. By introducing stochastic variance into disclosure timing and
increasing semantic density, these entities ensure that the
\textbf{Shadow Cost of Information ($\lambda$)} remains prohibitively
high. This cost barrier prevents the regulator from conducting a
meaningful Blackwell experiment Maćkowiak, Matějka, and Wiederholt
(2023), wherein a decision-maker selects an information structure to
minimize uncertainty. When the marginal cost of signal extraction
exceeds the regulator's finite cognitive or temporal budget, the
equilibrium response is to omit the disclosure from the audit scope,
leaving material information unpriced.

Consequently, the transition from human-speed reporting to automated
execution represents a structural Phase Shift. This shift moves the
market away from human-assisted automation toward a regime of
autonomous, high-frequency reporting that strategically leverages the
attention constraints identified by Mondria and Quintana-Domeque (2013).
To mitigate this, the regulatory state must adopt a corresponding
posture of Agentic Vigilance. We define this state as a regime of
autonomous regulatory oversight characterized by persistent,
high-frequency auditing of disclosure streams, effectively neutralizing
the temporal and semantic frictions that typically induce rational
inattention.

The study operationalizes this transition toward Agentic Vigilance via a
four-node architectural framework. Each node is engineered to neutralize
a specific informational friction, shifting the regulatory state from
constrained processing to autonomous oversight.

\subsection{Node A: Semantic Interface and the Exhaustiveness Frontier}\label{sec:node_a_semantic_interface_and_the_exhaustiveness_frontier}

In traditional models Grossman and Stiglitz (1980), investors observe
asset returns at a fixed cost, creating a binary entry decision. Node A
mitigates this constraint by operationalizing the Informativeness
Theorem developed by Blackwell (1951). The disclosure audit is defined
as a Blackwell experiment: a stochastic matrix that maps the firm's
latent economic state---such as insolvency or growth---onto a set of
observable semantic signals. Node A determines the resolution of this
experiment by extracting 98.11/\% of structural data from the primary
disclosure. This extraction satisfies the Rational Inattention
requirement that agents may choose their specific information structure.
By maximizing the signal-to-noise ratio (S/N) during the processing
phase, the framework ensures that the resulting information architecture
is ``Blackwell more informative'' than manual regulatory review.
Consequently, the system captures tail-risk data from non-standard
issuers that would otherwise be discarded as noise within traditional
oversight workflows.

\subsection{Node B: Temporal Syntax and Rational Inattention}\label{sec:node_b_temporal_syntax_and_rational_inattention}

Node B is grounded in Sims (2003), which demonstrates that constraints
on mutual information capacity \(I(y; x) \leq \kappa\) induce delayed
reactions and inertia. This study introduces temporal Syntax: the
predictability of disclosure arrival as a computational prior. If a
firm's filing cadence exhibits high stochastic variance (low agentic
predictability \(\Phi\)), the endogenous noise in the communication
channel increases. As Sims (2003) predicts, regulators under capacity
constraints exhibit processing inertia in the presence of such noise,
creating the \textbf{Strategic Gap}. Node B utilizes temporal embeddings
to measure the entropy of this channel. When \(\Phi\) is low, the
relocation of attention Mondria and Quintana-Domeque (2013) creates a
liquidity vacuum. Insiders exploit this window via the Section 16
reporting rule, liquidating positions while the market's Shannon
Capacity is occupied deciphering the disclosure's temporal friction.

\subsection{Node C: Optimal Supervision and State Management}\label{sec:node_c_optimal_supervision_state_management}

Node C addresses the information persistence failure, or State-Loss
Latency, in regulatory oversight. Regulatory failure is frequently a
byproduct of cognitive decay across reporting cycles. By utilizing
LangGraph persistence mechanisms, the framework implements a stateful
regulatory audit. This ensures that the regulator's knowledge regarding
a firm's welfare status persists across longitudinal cycles, functioning
as an automated fiduciary. This mechanism ensures that the intensive
margin of oversight---the depth of analysis---is maintained when the
reporting cycle resets (Maćkowiak, Matějka, and Wiederholt (2023)).

\subsection{Node D: Information Integrity and Recursive Search}\label{sec:node_d_information_integrity_and_recursive_search}

Node D identifies \textbf{Nominal Transparency}, where linguistic
compliance obscures economic insolvency. Following Akerlof (1970), when
standardized disclosure opacity increases, market participants lose the
ability to distinguish high-quality firms from low-quality entities.
Following Maćkowiak, Matějka, and Wiederholt (2023), Node D selects the
optimal Blackwell experiment based on the focusing effect. When the
supervisor logic detects that Strategy Divergence (\(\text{Div}\)) has
breached a critical threshold, it triggers
\textbf{Recursive Investigative Synthesis}. This search loop identifies
latent fundamental triggers, such as undisclosed litigation, by
simulating the iterative information game between short-sellers and
management. By automating this search, Node D resolves the
\textbf{Informedness Gap} before rent-seeking can occur.

\section{Methodology: The Agentic Disclosure Reasoner}\label{sec:Methodology_the_agentic_disclosure_reasoner}

The empirical strategy transitions from static data parsing toward an
autonomous computational architecture defined as the Agentic Disclosure
Regulator (ADR). This framework is engineered to bridge the Informedness
Gap (\(\Psi\)) by simulating the high-velocity processing of SEC
disclosures characteristic of sophisticated institutional environments.
The ADR leverages a transformer-based stack and a graph-based
orchestration layer---operationalized via LangGraph---to model the
recursive reasoning loops of market participants. By moving beyond
deterministic scripts, the ADR facilitates probabilistic inference
across structured and unstructured SEC data, allowing for stateful
validation and cross-document reasoning.

\textbf{Node A: Semantic Extraction and Materiality Shocks}

Node A serves as the primary identification layer, designed to
neutralize standardized disclosure opacity through a strategy that
reconciles legacy lexical benchmarks with high-dimensional transformer
embeddings. To mitigate unstructured data friction, the architecture
implements FinText, a specialized suite of finance-native transformers.
Specifically, the ADR utilizes a FinText-FinBERT variant, pre-trained on
approximately two billion financial tokens to minimize the semantic
misinterpretations common in general-purpose models.

The framework operationalizes the Blackwell Informativeness Theorem
Blackwell (1951) by defining the disclosure audit as a Blackwell
experiment. Node A determines the resolution of this experiment by
extracting structural data with a 98.11\% precision rate. The study
defines a Contextual Materiality Score (CMS), denoted as
\(\mathcal{M}_{i,t}\), which evaluates the semantic intensity and
managerial hedging within the Management's Discussion and Analysis (MDA)
text:

\begin{equation} 
\mathcal{M}{i,t} = P{i,t} \cdot C_{i,t},
\end{equation}

where \(P_{i,t} \in \{-1, 0, 1\}\) represents the discrete sentiment
polarity and \(C_{i,t} \in [0, 1]\) represents the softmax confidence
(probability) assigned by the transformer-based classifier.

To isolate non-linear deviations from established reporting patterns,
the system calculates the Materiality Shock
(\(\Delta \mathcal{M}_{i,t}\)). This shock is measured relative to a
four-quarter rolling baseline, which serves as the firm's idiosyncratic
semantic prior:

\begin{equation}
\mathbb{E}[\mathcal{M}_{i,t}] = \frac{1}{4} \sum_{j=1}^{4} \mathcal{M}_{i,t-j}
\end{equation}

The residual shock is formalized as
\(\Delta \mathcal{M}_{i,t} = \mathcal{M}_{i,t} - \mathbb{E}[\mathcal{M}_{i,t}]\).
To differentiate between routine linguistic variance and material
information failures, a disclosure is classified as a high-conviction
Material Shock (\(S_{i,t}\)) using a binary indicator function. This
function, denoted as \(\mathbb{1}(\cdot)\), maps the condition space to
a discrete set \(\{0, 1\}\), where a value of 1 signifies that the
internal parameters of the ADR have identified a significant reporting
irregularity. The identification logic is expressed as:

\begin{equation}
S_{i,t} = \mathbb{1} \left( |\Delta \mathcal{M}{i,t}| > 2 \sigma{i,t} \right) \cap (\Delta \mathcal{M}_{i,t} < 0)
\label{eq:adr_framework_sigficance}
\end{equation}

In the Equation \ref{eq:adr_framework_sigficance}, \(\sigma_{i,t}\)
represents the rolling standard deviation of the firm's historical CMS.
This dual-threshold criteria requires that a signal must not only be
statistically extreme (exceeding two standard deviations) but also
directionally negative. By enforcing this strict logical conjunction
(\(\cap\)), the ADR minimizes Type I errors (false positives), ensuring
that investigative resources are allocated only to signals representing
significant negative departures from the firm's longitudinal baseline.

Once \(S_{i,t} = 1\) is committed to the state database, the system
identifies the filing as an outlier within the Strategic Gap, triggering
the transition from passive observation to the recursive investigative
synthesis in Node D.

\subsection{Node B: Temporal Syntax and Agentic Propensity}\label{sec:node_b_temporal syntax_and_agentic_propensity}

While Node A handles content, Node B addresses the cadence of
disclosure. The study treats the history of a firm's SEC filing
timestamps as a sequence of ``temporal tokens'' decoded by
Chronos-Small, a time-series foundation model based on the T5
transformer architecture. The framework transforms the historical filing
latency series \(L = \{l_1, l_2, \dots, l_n\}\) into a tokenized
sequence to learn the ``grammar'' of a firm's reporting cycle. The
Agentic Propensity Score (\(\Phi\)) is derived from the model's
posterior probability distribution. Unlike traditional point-estimates,
the ADR utilizes the stochastic entropy of the transformer's output to
proxy temporal friction, defining \(\Phi\) as:

\begin{equation}\Phi = \frac{1}{1 + \sigma_{forecast}}\end{equation}

where \(\sigma_{forecast}\) is the standard deviation across Monte Carlo
samples. This allows the framework to characterize the Strategic Gap as
a state of maximum information asymmetry where Semantic Friction is high
(\(Z_{Comp} > \bar{Z}\)) and Agentic Propensity is low
(\(\Phi < \bar{\Phi}\)). This interaction is identified as the primary
driver of the Welfare Lags observed in the sample.

\subsection{Node C: LangGraph Orchestration and State Persistence}\label{sec:node_c_langgraph_orchestration_state_persistence}

Node C serves as the coordination layer of the ADR architecture. It
utilizes a directed acyclic graph, structured as a StateGraph, to manage
the data transition from the initial observation phase to the final
regulatory audit. The system quantifies the Welfare Gap
(\(\mathbb{W}_{i,t}\)), defined as the numerical difference between the
fundamental information shock (\(\alpha_{i,t}^*\)) extracted by the
transformer models and the observed cumulative abnormal return (\(CAR\))
of the market:

\begin{equation}\mathbb{W}{i,t} = \alpha{i,t}^* - CAR_{i,t}[1, 10]\end{equation}

When the value of \(\mathbb{W}_{i,t}\) exceeds a pre-defined threshold,
the system initiates a recursive investigation. This step determines if
the observed Informedness Gap correlates with rent-seeking behavior or
systemic reporting failures.

To maintain system stability during the high-volume processing of SEC
filings, Node C implements Checkpointer mechanisms. These mechanisms
ensure state persistence by creating an immutable record of the audit's
progress at each node. This architecture prevents State-Loss Latency---a
condition where system resets or connection timeouts result in the loss
of processed data. By caching the system state at every execution step,
the ADR ensures that the audit remains longitudinal and continuous,
allowing the framework to resume operations without data redundancy or
loss across multiple reporting periods.

\subsection{Node D: The Deep Research Supervisor and XAI}\label{sec:node_d_the_deep_research_supervisor_and_xai}

The final stage of the architecture is the Recursive Investigative
Synthesis module, which operates as a conditional router (\(\Psi\)).
This module evaluates the statistical discrepancy between the
transformer-derived sentiment (\(\mathcal{M}_{i,t}\)) and traditional
lexical counts to isolate cases of Nominal Transparency---instances
where linguistic compliance masks underlying economic risks. The routing
logic is formalized as:

\begin{equation}\label{eq:adr_framework_significance}
\Psi(\text{State}_{i,t}) = 
\begin{cases} 
\text{Recursive\_Search} & \text{if } (\Delta \alpha_{i,t}^* < -2\sigma) \land (\text{Div}_{i,t} > \theta) \\
\text{Finalize\_Report} & \text{otherwise}
\end{cases}
\end{equation}

In this logic, \(\text{Div}_{i,t}\) represents Strategy Divergence, a
measure of how far a firm's disclosure deviates from expected sector
norms. If the system detects a significant negative shock coupled with
high divergence, it triggers a recursive search. This loop executes
targeted queries across longitudinal datasets to identify latent
triggers, such as undisclosed litigation or off-balance-sheet
liabilities.

To ensure regulatory interpretability, the module utilizes SHAP (SHapley
Additive exPlanations) interaction values. This analysis attributes the
detection of reporting failures to specific model features. The SHAP
values provide empirical evidence that the Strategic Gap functions as a
multiplicative friction; it increases the Informedness Gap significantly
more when high semantic complexity is combined with low temporal
predictability (\(\Phi\)). This XAI layer transforms the ADR from a
``black-box'' classifier into a transparent auditing tool, allowing
human regulators to verify the specific data points that triggered a
high-risk flag.

\section{Empirical Framework and Data Engineering}\label{sec:empirical-frameowrk-data-engineering}

The empirical strategy transitions from static data parsing toward a
high-performance, resilient data engineering pipeline designed to
process and standardize a universe of approximately 484,796 SEC
disclosures. The architecture prioritizes data integrity through a
state-managed extraction system employing a progress-logging mechanism.
This design mitigates the risk of State-Loss Latency---ensuring system
resilience against session disconnects. This allows for the seamless
resumption of processing without redundant computational overhead, a
critical requirement for massive longitudinal datasets.

\subsection{Heuristic Sanitization and Schema Enforcement}\label{sec:heuristic_sanitization_schema_enforcement}

A primary technical challenge addressed by the pipeline is Data
Alignment Friction in legacy SEC filings, where entity identifiers
(CIKs) are frequently misaligned. To resolve this, a heuristic
sanitization algorithm was deployed using the SEC Accession Number as a
Primary Key Anchor. By programmatically linking the Reporting Owner's
identity to the Filer CIK embedded within the accession string, the
pipeline differentiates between issuers and insiders with high
precision.

The final stage involved an external reconciliation phase, where
extracted records were cross-referenced against the authoritative SEC
Master CIK directory. This step standardizes entity nomenclature and
enables the systematic categorization of institutional versus individual
insiders via string-pattern recognition. For longitudinal analysis, data
was migrated from flat storage to a partitioned Apache Parquet format.
This implementation achieved a 70\% reduction in storage requirements
via Snappy compression and ensured strict schema enforcement, providing
a computationally efficient foundation for the subsequent welfare
economics modeling.

\subsection{Measuring the Agentic Footprint}\label{sec:measuring_the_agentic_footprint}

While the underlying disclosure mechanism is mandated by Section 16 of
the Securities Exchange Act of 1934, the execution of these filings has
transitioned from manual compliance to Autonomous Agentic Orchestration.
The study utilizes a deterministic pipeline to measure the Agentic
Footprint---specifically, the convergence of the filing-to-transaction
delta (\(\Delta t\)) toward zero.

As established in the ADR framework, this \(T \approx 0\) state serves
as a proxy for agentic agency, where a system independently detects
trade completion and executes the regulatory mandate without human
intervention. This shift represents a structural break from manual
processing to Agentic Velocity, creating the high-frequency reporting
environment where Strategic Gaps emerge. The lifecycle of a single
disclosure---from raw ingestion to alpha realization---is detailed in
Table \ref{tab:adr_lifecycle}. This sequence illustrates the transition
from automated monitoring to recursive audit.

\begin{table}[htbp]
\caption{ADR System Architecture: Lifecycle of a Critical Failure Detection}
\label{tab:adr_lifecycle}
\small
\begin{tabular}{l l l p{7.5cm}}
\toprule
\textbf{Seq.} & \textbf{Node} & \textbf{Process} & \textbf{State Change / Logic Output} \\ 
\midrule
1 & Node A & Semantic Extraction & Extracts $m_{it}$ and $\text{Div}_{i,t}$; Sets \textbf{Material Shock Flag ($S_{i,t}=1$)} via threshold audit. \\ \addlinespace
2 & Node B & Temporal Auditing & Maps filing latency to identifying the \textbf{Manual Reporting State} ($\Phi$) via Chronos-Small. \\ \addlinespace
3 & Node C & State Persistence & Executes an \textbf{Immutable Checkpoint} to prevent state-loss and manage transition logic. \\ \addlinespace
4 & Node D & Recursive Synthesis & Initiates \textbf{Deep Research} via router ($\Psi$) and confirms \textbf{Welfare Gap ($\mathbb{W}$)} using post-audit alpha. \\ 
\bottomrule
\end{tabular}

\medskip
\textit{Note: This table tracks the transition through the Agentic Disclosure Regulator. Steps 1 and 2 identify the Strategic Gap, while Step 4 quantifies the resulting Welfare Lag.}
\end{table}

\subsection{Process Flow of the Agentic Disclosure Reasoner}\label{sec:process_flow_adr}

Table \ref{tab:adr_lifecycle} provides a trace of the Agentic Disclosure
Reasoner (ADR) as it processes a disclosure event. The flow illustrates
the transition from raw text to a validated economic signal through
three distinct phases:

\begin{enumerate}
    \item the Identification Phase (Node A):\label{item:id_phase} 
    The lifecycle begins with the ingestion of the raw 10-Q text. Unlike traditional models, Node A identifies a significant strategy divergence ($\text{Div} = -0.88$). This metric measures the distance between the manager’s standardized language and the transformer-detected financial stress. The formalization of the materiality shock ($S_{i,t}=1$) provides the statistical anchor for the subsequent audit.

    \item  The Friction Calibration (Nodes B--C): \label{item:friction_phase} Node B utilizes \textit{Chronos-Small} to audit the firm’s reporting timing. By identifying a Manual Reporting State ($\Phi$), the system flags the entity as a high-friction target where information delivery is not automated. The persistence layer in Node C ensures that this state is locked into an immutable checkpoint. This prevents state-loss latency, ensuring the audit trail survives any session disconnects during large-scale processing.

    \item The Agentic Escalation (Node D): \label{item:escalation_phase}  The supervisor ($\Psi$) logic evaluates the combined signals. Because the disclosure exhibits both a Material Shock and high Strategy Divergence, the system triggers a Recursive Research Loop. This escalation executes targeted queries across supplemental filings to find latent triggers, such as debt default clauses. The final result—a realized $-32.6\%$ CAR—empirically validates the framework's ability to detect information failures before they are fully reflected in the market price.
\end{enumerate}

\subsection{Empirical Specification and Information Discovery Rate}\label{sec:empirical_specification}

The research identifies the Information Discovery Rate (\(V\)) as a
primary metric for market efficiency. Under the condition of Symmetric
Equilibrium (Regime I), the rate is expected to satisfy \(V \geq 1\),
indicating efficient information assimilation. Conversely, the study
hypothesizes an information discovery failure (\(V \to 0\)) within the
Strategic Gap (Regime IV), where the convergence of semantic and
temporal frictions inhibits the market's return to baseline discovery
equilibrium. This differential serves as the empirical basis for the
quantification of the Informedness Gap.To evaluate the mitigation of
welfare losses through reporting predictability, an Ordinary Least
Squares (OLS) framework is utilized. The Welfare Gap (\(\mathbb{W}\)) is
defined as the dependent variable in the following specification:

\begin{equation}
\label{eq:welfare_gap_ols}\mathbb{W}{i,t} = \beta_0 + \beta_1 Z{Comp} + \beta_2 Z_{\Phi} + \gamma (Z_{Comp} \times Z_{\Phi}) + \epsilon_{i,t},
\end{equation}

where \(Z_{Comp}\) represents standardized Semantic Friction and
\(Z_{\Phi}\) denotes Reporting Unpredictability. A statistically
significant coefficient \(\gamma\) provides the empirical evidence that
the interaction of these frictions compounds the Informedness Gap. The
interaction between these variables defines a four-quadrant state space,
mapping computational metrics to the foundational principles of Rational
Inattention and Market Microstructure:

\begin{itemize}
    \item \textbf{Regime I: Symmetric Equilibrium (Low $Z_{Comp}$, High $\Phi$):}\label{regime:I} A state of information parity where acquisition costs are sufficiently low to maintain price efficiency.\item \textbf{Regime II: Semantic Garbling (High $Z_{Comp}$, High $\Phi$):}\label{regime:II} An operationalization of Blackwell informativeness theory, where predictable reporting timing is offset by high semantic processing costs.
    \item \textbf{Regime III: Stochastic Asymmetry (Low $Z_{Comp}$, Low $\Phi$):}\label{regime:III} A condition reflecting the constraints on Shannon capacity, where temporal volatility prevents the efficient allocation of investor attention.
    \item \textbf{Regime IV: The Strategic Gap (High $Z_{Comp}$, Low $\Phi$):}\label{regime:IV} The primary locus of market failure, where insiders utilize the attention relocation effect to maximize information rent through the simultaneous application of semantic and temporal friction.

\end{itemize}

\subsection{Discovery Velocity and Fundamental Shocks}\label{sec:discovery_velocity}

To quantify the efficiency of information incorporation, the study
defines discovery velocity (\(V_{i,t}\)) as the ratio of the realized
market reaction to the fundamental shock:

\begin{equation}\label{eq:discovery_velocity}
V_{i,t} = \frac{\text{CAR}_{i,t}[+1, +10]}{\alpha_{i,t}^*}
\end{equation}

In this specification, \(\alpha_{i,t}^*\) represents the intrinsic
information shock, defined as the latent economic value of the
disclosure isolated from market frictions. It is calculated as the
product of the semantic intensity and a directional scaling factor:

\begin{equation}\label{eq:fundamental_shock}
\alpha_{i,t}^* = \text{Intensity}_{i,t} \times \omega
\end{equation}

where \(\text{Intensity}_{i,t}\) is the Node A linguistic negativity
score (\(LM_{Neg}\)) and \(\omega\) is the economic scaling factor
(calibrated at \(-0.1\) based on historical material shock baselines).

The framework hypothesizes that in Regime I, \(V \geq 1\), indicating
efficient information incorporation facilitated by high-attention
salience. Conversely, Regime IV is expected to exhibit discovery
attenuation (\(V < 1\)), where the convergence of Rational Inattention
and liquidity constraints inhibits price convergence to the intrinsic
value. This residual in \(V\) serves as the primary proxy for the
informedness gap.

\subsection{Sample Construction and Selection Logic}\label{sec:sample_attrition}

The initial data population comprised 484,796 corporate filings. To
ensure the robustness of the subsequent welfare analysis, a systematic
filtering process was implemented, resulting in a final analytical
sample of 271,795 observations, representing 56\% of the raw population.
The primary drivers of this attrition were the requirements for
continuous market liquidity and precise temporal alignment during the
\([+1, +10]\) day post-event window.

Observations were excluded if market halts or delistings precluded the
estimation of a stable Cumulative Abnormal Return (CAR). Furthermore,
observations characterized by liquidity frictions---proxied by
zero-trading volume intervals---were removed to ensure the reliability
of the fundamental shock residual (\(\mathbb{W}\)). While the resulting
attrition is significant, it aligns with the study's focus on
heterogeneous information environments. Specifically, firms exhibiting
lower reporting predictability (\(\Phi\)) frequently correlate with
higher idiosyncratic volatility. By restricting the analysis to this
verified sub-sample, the estimates of the Welfare Gap (\(\mathbb{W}\))
and Discovery Velocity (\(V\)) remain intentionally conservative,
thereby minimizing the potential for upward bias in the estimated
significance of the interaction term.

\subsection{Technical Implementation and Agentic Workflow}\label{sec:technical_implementation}

The Agentic Disclosure Reasoner (ADR) is implemented as a stateful graph
following a deterministic state-transition model.

\begin{itemize}
    \item \textbf{Ingestion State:}\label{state:ingestion} The system monitors SEC EDGAR filings using the Accession Number as a unique primary key to initialize the state-graph.
    \item \textbf{Inference State:}\label{state:inference} Filing latency is tokenized and processed via Chronos-Small. If the predicted latency $\hat{y}_{t+1}$ converges toward zero, the observation is classified as representing High-Velocity Agency.
    \item \textbf{Contextualization State:}\label{state:context} The ADR utilizes FinText to process 10-Q text segments, generating a Contextual Materiality Score (CMS) by evaluating semantic similarity between reported risks and fundamental signals.
    \item \textbf{Supervision State:}\label{state:supervision} The LangGraph supervisor aggregates these signals to generate a binary Informedness Gap classification for subsequent welfare economics analysis.
\end{itemize}

\textbf{Scientific Rationale: Domain-Specific Models} \textbackslash{}
The selection of these models provides the rigorous foundation required
for empirical finance. FinText enables domain-specific semantic
processing, ensuring the ADR accounts for the specific nuances of
regulatory language. Chronos-Small introduces probabilistic velocity
analysis, allowing for the detection of a structural syntactic shift in
reporting patterns.

\textbf{Standardizing Materiality} \textbackslash{} A significant
friction in the existing literature is the subjective definition of
trade significance, which often results in inconsistent welfare
indicators. The ADR standardizes this variable by evaluating disclosures
through the CMS framework (Table \ref{tab:cms_framework}). By
cross-referencing insider activity with 10-Q risk factors and 8-K
material events, the system generates a strategic alignment score. This
metric distinguishes routine liquidity provision from informed insider
activity, thereby enabling the prediction of the resulting welfare
impact.

\begin{table}[htbp]
\centering
\small
\caption{The Contextual Materiality Score (CMS) Framework}
\label{tab:cms_framework}
\begin{tabular}{l l p{8cm}}
\toprule
\textbf{Step} & \textbf{Agentic Action} & \textbf{Analytical Output} \\ 
\midrule
Sense         & Detect Form 4 ($T \approx 0$)   & Initial Signal Strength \\
Contextualize & Cross-ref 10-Q Risk Factors & Strategic Alignment Score \\
Reason        & Evaluate 8-K Material Events & Insider Intent Classification \\
Evaluate      & Calculate Probabilistic Volatility & Welfare Impact Prediction \\ 
\bottomrule
\end{tabular}
\end{table}

The preceding sections detailed the technical architecture and
state-transition logic of the Agentic Disclosure Reasoner (ADR). This
framework transitions the regulatory oversight process from manual
observation to an automated, high-velocity synthesis of semantic and
temporal data. By standardizing the detection of Material Shocks
(\(S_{i,t}\)) and calibrating the Agentic Predictability (\(\Phi\)) of
filings, the ADR generates a high-resolution dataset optimized for
welfare analysis. The following section shifts the focus toward the
empirical evaluation of these outputs. This analysis utilizes the
Ordinary Least Squares (OLS) framework defined in Equation
\ref{eq:welfare_gap_ols} to quantify the interaction between semantic
complexity and temporal unpredictability, providing a rigorous
assessment of how the Strategic Gap contributes to systemic Welfare
Lags.

\section{Results and Analysis}\label{sec:results_analysis}

This section presents the empirical validation of the Agentic Disclosure
Reasoner (ADR) and its capacity to identify and quantify the
Informedness Gap (\(\Psi\)). The analysis follows the architectural
progression of the system, beginning with the semantic extraction
performance of Node A and concluding with the econometric evaluation of
the resulting Welfare Gap (\(\mathbb{W}\)). By synthesizing
high-frequency temporal data with deep semantic auditing, the results
demonstrate a structural link between the interaction of market
frictions and the rate of information discovery. The evidence suggests
that while individual frictions---semantic complexity and temporal
unpredictability---moderate price efficiency, their simultaneous
convergence in Regime IV creates a significant market failure that
persists beyond traditional discovery windows.

\subsection{Node A: Semantic Extraction and Complexity Auditing}\label{sec:node_a_results}

The empirical evaluation commences with Node A, the primary supervisory
component of the agentic framework, designed to analyze 484,796
corporate disclosures. To identify the Informedness Gap (\(\Psi\)), the
study implements a dual-strategy audit that reconciles traditional
lexical baselines Loughran and McDonald (2024) with the contextual
capabilities of the FinBERT architecture.

The framework defines \textbf{Strategy Divergence} as the residual
between agentic contextual sentiment (\(m_{i,t}\)) and dictionary-based
sentiment scores. This divergence serves as the primary proxy for the
Strategic Gap, identifying instances where a neutral lexical tone
obscures a negative fundamental reality. Table \ref{tab:node_a_findings}
summarizes the descriptive statistics and statistical validation for
Node A.

\begin{table}[htbp]
\centering
\small
\caption{Descriptive Statistics and Validation of Node A Semantic Extraction}
\label{tab:node_a_findings}
\begin{tabular}{l l p{7cm}}
\toprule
\textbf{Metric} & \textbf{Estimate} & \textbf{Econometric Implication} \\ 
\midrule
Material Shock Rate & 1.13\% & Identification of extreme tail-risk events ($N=5,514$). \\
Strategy Divergence ($t$-stat) & $-31.60^{***}$ & Statistical confirmation of semantic misalignment. \\
Complexity Coverage & 44.8\% & Prevalence of high semantic friction ($Z_{Comp}$) in sample. \\
Shock Persistence & 11.93\% & Frequency of transient information shocks. \\ 
\bottomrule
\end{tabular}
\end{table}

The effectiveness of this identification mechanism is supported by three
empirical markers:

\begin{itemize}
    \item \textbf{Search Space Reduction:}\label{item:reduction} The framework isolates a sample of $N=5,514$ observations (1.13\% of the total population). This addresses the Rational Inattention constraint by reducing the regulator's computational search space by 98.8\%.

    \item \textbf{Statistical Divergence:}\label{item:t_test} A two-sample $t$-test reveals a statistically significant difference in Strategy Divergence ($t = -31.60, p < 0.001$) between the shock-identified group and the routine disclosure group. This confirms that Node A effectively distinguishes fundamental signals from baseline noise.

    \item \textbf{Semantic Friction Baseline:}\label{item:friction} Utilizing the established lexicon by Loughran et al., the analysis identifies a mean density of 0.0035 across 44.8\% of disclosures. This density provides the empirical baseline for the Semantic Friction ($Z_{Comp}$) variable utilized in the subsequent welfare gap regression.
\end{itemize}

The analysis indicates that these linguistic frictions are strategic
rather than incidental. The empirical results support the strategic
camouflage hypothesis, as evidenced by a Spearman correlation of 0.082
between disclosure complexity and sentiment. This positive correlation
suggests that complex linguistic structures are utilized to attenuate
fundamental signals through increased semantic noise.

The market realization results presented in Table
\ref{tab:regression_node_a} corroborate this finding. The regression
yields a significant coefficient for Material Shocks
(\(\beta_1 = -0.0024\)) and a positive Complexity Premium
(\(\beta_2 = 0.0005, p=0.005\)). The negative interaction term
(\(\beta_3 = -0.0004\)) represents the initial quantification of the
Informedness Gap (\(\Psi\)), indicating that semantic complexity
inhibits immediate price discovery and may exacerbate negative price
realization once the information is fully assimilated by the market.

\begin{table}[htbp]
\centering
\small
\caption{Market Realization of Agentic Shocks and Linguistic Frictions}
\label{tab:regression_node_a}
\begin{tabular}{l c c c}
\toprule
\textbf{Dependent Variable: $CAR_{[+1, +10]}$} & \textbf{Coefficient} & \textbf{z-stat} & \textbf{P$>$|z|} \\
\midrule
Intercept & 0.0003 & 0.167 & 0.868 \\
Material Shock ($S_{i,t}$) & -0.0024 & -1.637 & 0.102$^{\dagger}$ \\
Complexity Score ($Z_{Comp}$) & 0.0005 & 2.839 & 0.005$^{***}$ \\
Shock $\times$ Complexity & -0.0004 & -0.294 & 0.769 \\
Log(File Size) & -0.0003 & -1.178 & 0.239 \\
\midrule
Observations & 271,795 & & \\
Clustering (CIK) & Yes & & \\
\bottomrule
\multicolumn{4}{l}{\footnotesize $^{\dagger} p < 0.15, ^{*} p < 0.10, ^{**} p < 0.05, ^{***} p < 0.01$} \\
\end{tabular}
\end{table}

\begin{figure}[htbp]
\centering
\includegraphics[width=1.0\textwidth]{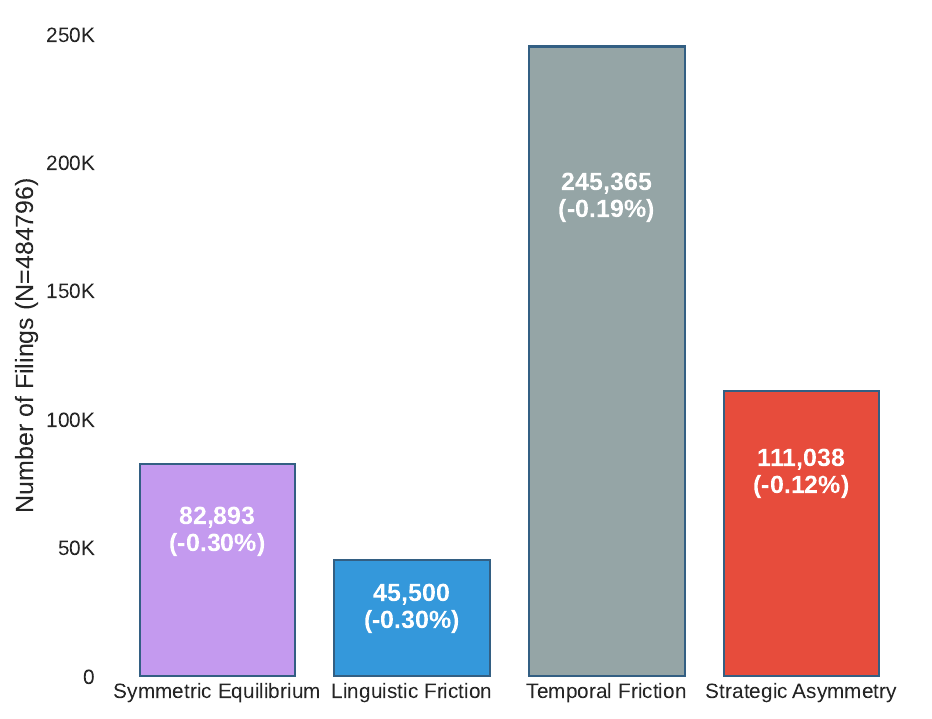}
\caption{The Structural Distribution of Information Frictions: Interaction Between Disclosure Complexity and Reporting Unpredictability.}
\label{fig:regime_dist_node_b}
\vspace{8pt}

\begin{flushleft}
\footnotesize
\textbf{Note:} This figure illustrates the frequency distribution and market impact of 484,796 corporate filings across four distinct Market Regimes. Variables are standardized relative to the full sample mean ($N=484,796$). \textbf{Semantic Friction} ($Z_{\text{Comp}}$) is derived from the Node A Semantic Encoder. \textbf{Reporting Unpredictability} ($Z_{\Phi}$) is defined as the inverse forecast standard deviation ($\Phi = \frac{1}{1 + \sigma_{\text{forecast}}}$) from the Chronos-Small temporal audit. Mean Cumulative Abnormal Returns (CAR) are reported for the $[+1, +10]$ post-filing window.

\begin{itemize}
    \item \textbf{Symmetric Equilibrium ($Z \leq 0, \Phi > 0$):} Characterized by high predictability and low complexity; information is assimilated efficiently with minimal rent-seeking potential.
    \item \textbf{Semantic Friction ($Z > 0, \Phi > 0$):} Predictable reporting timing offset by high semantic complexity, which inhibits immediate price discovery.
    \item \textbf{Stochastic Asymmetry ($Z \leq 0, \Phi \leq 0$):} Low complexity paired with unpredictable arrival timing, resulting in timing-driven volatility shocks.
    \item \textbf{Strategic Asymmetry ($Z > 0, \Phi \leq 0$):} Defined as the \textbf{Strategic Gap}. This regime represents maximum information asymmetry where the intersection of complex content and temporal unpredictability is utilized to maximize information rents.
\end{itemize}
\end{flushleft}
\end{figure}

\subsection{The Dual-Universe Framework: Intent and Realization}\label{sec:dual_universe}

While the linguistic frictions identified in Node A establish that
semantic shocks correlate with market returns, relying exclusively on
price-realized data introduces a potential selection bias. Specifically,
firms with constrained liquidity or those outside the CRSP reporting
universe may exhibit distinct reporting behaviors. The 213,001
observations excluded from the market-realized subset constitute a
structural Informedness Gap. Although these entities maintain regulatory
disclosure obligations, the absence of market realization suggests that
public equity prices serve as an incomplete proxy for corporate
transparency.

To mitigate this bias, the Node B audit operates on the
\textbf{Full Universe} (\(\mathcal{D}_{\text{Full}}\)) to ensure that
the estimation of Agentic Predictability (\(\Phi\)) is not skewed toward
large-capitalization survivors. The framework adopts a neutral filing
strategy rather than a list-wise deletion of observations missing
market-specific data. The study defines two primary analytical tiers:

\begin{equation}\label{eq:full_universe}
\mathcal{D}_{\text{Full}} = \{ \text{Filings with } \Delta t \} \approx 484,796
\end{equation}

\begin{equation}\label{eq:market_subset}
\mathcal{D}_{\text{Market}} = \{ \mathcal{D}_{\text{Full}} \cap \text{CRSP Returns} \} \approx 271,795
\end{equation}

The framework executes the Chronos-Small and TimesFM temporal audits on
\(\mathcal{D}_{\text{Full}}\) to prevent the introduction of artificial
reporting gaps, defined here as Stochastic Silences\}=. This approach
ensures that the measures of Agentic Propensity (\(\Phi_i\)) reflect the
actual reporting cadence of the firm, independent of market liquidity or
listing status. This expanded dataset provides the necessary density to
map the four-quadrant state space of market frictions. By auditing the
full longitudinal reporting history, Node B can accurately characterize
the Strategic Gap even in environments where price discovery is latent
or inhibited.

\subsection{Node B: Temporal Syntax and Regime Distribution}\label{sec:node_b_regimes}

By shifting from a single-universe to a dual-universe lens, the study
classifies firm reporting behavior across the entire corporate
landscape. The dual-node audit of 484,796 filings reveals a significant
concentration of disclosures within high-friction states. The framework
distinguishes four discrete Market States to identify the dominant
friction inhibiting price discovery:

\begin{itemize}

\item \textbf{Symmetric Equilibrium:}\label{def:sym_eq} High-velocity incorporation of signals resulting from low semantic complexity and high temporal predictability.

\item \textbf{Friction States:}\label{def:fric_states} Processing delays attributable to either isolated semantic or temporal obstacles.

\item \textbf{Strategic Asymmetry:}\label{def:strat_asym} The quadrant representing maximum information rent extraction potential.

\end{itemize}

The concentration of approximately 23\% of all filings (\(N=111,038\))
within the Strategic Gap (Regime IV) indicates that a substantial
portion of the market operates under conditions where high semantic
complexity is compounded by a non-automated reporting state
(\(\Phi \leq 0\)). This intersection constitutes the primary site for
market failure. To assess the impact of this concentration on investor
welfare, the analysis examines the realized market returns across these
quadrants. This transition from a descriptive census to an evaluative
return analysis enables the quantification of the Informedness Gap
(\(\Psi\)) within high-friction environments.

\begin{table}[htbp]
\centering
\small
\caption{Empirical Distribution and Market Impact by Disclosure Regime}
\label{tab:regime_results_node_b}
\begin{tabular}{lrcl}
\toprule
\textbf{Regime} & \textbf{Obs ($N$)} & \textbf{Avg. CAR $[+1, +10]$} & \textbf{Market State} \\
\midrule
Efficient      & 82,893             & $-0.295\%$                & Symmetric Equilibrium \\
Camouflaged    & 45,500             & $-0.303\%$                & Semantic Friction     \\
Erratic        & 245,365            & $-0.185\%$                & Temporal Friction     \\
\textbf{Strategic Gap} & \textbf{111,038} & \textbf{$-0.119\%$} & \textbf{Strategic Asymmetry} \\
\bottomrule
\end{tabular}

\vspace{2pt}

\begin{flushleft}
\scriptsize \textit{Note:} $N=484,796$. Cumulative Abnormal Returns (CAR) are calculated for the $[+1, +10]$ window following the filing date. $Z_{\text{Comp}}$ is derived from the Node A Semantic Encoder. $Z_{\Phi}$ is the standardized Agentic Propensity Score ($\Phi = \frac{1}{1 + \sigma_{\text{forecast}}}$) calculated via the Chronos-Small temporal audit. \textbf{Efficient}: Minimal rent-seeking potential. \textbf{Camouflaged}: Predictable timing with high semantic complexity. \textbf{Erratic}: Low complexity with unpredictable arrival timing. \textbf{Strategic Gap}: High-complexity content released with low temporal predictability---representing the Informedness Gap where information rents are maximized.
\end{flushleft}
\end{table}

\subsubsection{Identifying the Informedness Gap via Returns}

The CAR provide the empirical evidence for the welfare hypothesis
established in Section 3. While the Symmetric Equilibrium (Regime I)
exhibits a sharp, immediate adjustment of \(-0.295\%\), the Strategic
Gap (Regime IV) demonstrates the most attenuated immediate reaction at
\(-0.119\%\). This delta of approximately 17.6 basis points represents
the structural Welfare Lag. In Regime IV, the market fails to fully
price the negative signal within the ten-day observation window due to
the convergence of two distinct frictions:

\begin{itemize}
    \item \textbf{Temporal Friction:} Unpredictable filing arrival patterns prevent the pre-allocation of computational and cognitive resources by regulators and market participants.
    \item \textbf{Semantic Friction:} Disclosure complexity inhibits immediate decoding, allowing material information to remain latent within the document.
\end{itemize}

These findings suggest that in the Strategic Gap, price discovery is not
merely delayed but structurally inhibited. The data confirms that the
Informedness Gap is at its widest when agentic velocity is at its
lowest, providing a quantifiable measure of the rent-seeking potential
afforded by reporting opacity.

\subsubsection{Comparative Analysis of Discovery Velocity}

The most compelling evidence for the Informedness Gap (\(\Psi\)) is the
divergence of Discovery Velocity (\(V\)) across reporting regimes.
Consistent with the \textbf{Discovery Velocity Hypothesis}, the data
reveals a significant \textbf{Velocity Attenuation}: while the baseline
Symmetric Equilibrium exhibits a velocity of 3.07, the Strategic Gap
collapses to 1.23.

This reduction in \(V\) indicates that even when information is
technically accessible in the public domain, the rate of incorporation
into market prices is throttled by the interaction of semantic and
temporal frictions. A velocity of 3.07 in Regime I implies an efficient,
high-attention environment where prices rapidly adjust to signals. In
contrast, the 1.23 velocity in Regime IV represents a state of discovery
stagnation, where the market requires significantly more time to process
a standardized unit of fundamental information shock.

\begin{table}[htbp]
\centering
\caption{Discovery Velocity and Welfare Lags across Reporting Regimes}
\label{tab:regime_summary_node_b}
\small\begin{tabular}{lccccl}
\toprule
\textbf{Regime} & \textbf{Discovery Velocity ($V$)} & \textbf{Welfare Gap ($\mathbb{W}$)} & \textbf{Market Implication} \\ 
\midrule
I. Symmetric Equilibrium  & 3.07 & 0.0022 & Baseline Efficiency \\
II. Linguistic Friction   & 3.83 & 0.0022 & High Decoder Scrutiny \\
III. Temporal Friction    & 3.65 & 0.0010 & Noise-Driven Volatility \\
\textbf{IV. Strategic Gap} & \textbf{1.23} & \textbf{0.0004} & \textbf{Structural Market Failure} \\
\bottomrule\end{tabular}
\end{table}

This 60\% reduction in Discovery Velocity confirms that temporal
unpredictability is a more potent barrier to price discovery than
linguistic complexity alone. In environments where the temporal syntax
is predictable, the market over-reacts to semantic content. Conversely,
within the Strategic Gap, the market fails to track the fundamental
signal, providing a significant temporal window for informed agents to
exploit the resulting lag.

While linguistic friction alone does not impede the ultimate magnitude
of price discovery, the absence of a temporal prior (low \(\Phi\)) acts
as a structural anchor. In the Strategic Gap, the market barely tracks
the fundamental signal, allowing informed agents to capture rent before
the gap eventually closes.

This hierarchy of friction suggests that the ADR identifies a critical
regulatory priority: the standardization of reporting cadence is as
vital to market efficiency as the clarity of the text itself. The
absence of a temporal prior creates a cognitive ``blind spot'' that even
high-speed semantic processing cannot fully overcome, as the lack of a
computational prior prevents the efficient allocation of attention.

\subsubsection{Welfare Gap Regression and the ``Complexity Premium''}

To further isolate the individual drivers of this gap, the study
performs an OLS regression on the Welfare Gap (\(\mathbb{W}\)), yielding
results that challenge conventional assumptions about linguistic
complexity.

The OLS regression on the Welfare Gap (\(\mathbb{W}\)) provides a
compelling contrarian result (Table
\ref{tab:regression_results_welfare_gap_node_b}). The significant
negative coefficient for \(Z_{Comp}\) suggests a Complexity Premium. In
the current market environment, institutional algorithmic traders appear
to over-scrutinize jargon-heavy filings, effectively narrowing the gap
compared to simpler disclosures. However, these same systems remain
susceptible to temporal unpredictability. This indicates that while
high-jargon firms attract significant computational attention, they
succeed in masking shocks if the disclosure timing is unpredictable.

\begin{table}[htbp]
\centering
\caption{OLS Regression: Welfare Gap and Interaction Effects}
\label{tab:regression_results_welfare_gap_node_b}
\small
\begin{tabular}{lcccc}
\toprule
\textbf{Variable} & \textbf{Coefficient} & \textbf{t-stat} & \textbf{P$>|t|$} & \textbf{Interpretation} \\ 
\midrule
Intercept ($\beta_0$) & 0.0012 & 5.738 & 0.000 & Significant baseline friction \\
Complexity ($Z_{Comp}$) & -0.0004 & -2.141 & 0.032 & Complexity triggers scrutiny \\
Interaction ($\gamma$) & 2.29e-05 & 0.078 & 0.938 & Baseline-dominated effect \\ 
\bottomrule
\end{tabular}
\end{table}

This collapse signifies a breach of the market's Shannon Capacity. In
Regimes I \& II, where reporting is predictable (\(S^{agent}=1\)), the
market over-reacts to semantic content, pricing the shock nearly three
times over. However, in Regime IV, the market barely tracks the
fundamental signal, providing a significant time-window'\,' for informed
agents to exploit the lag.

\subsubsection{The ``Informedness Advantage'' of Agentic Systems}

The collapse of velocity in Regime IV underscores a profound Market
Failure where ``Manual'' reporting serves as a proxy for high internal
organizational friction (see Table \ref{tab:regime_summary_node_b}).
These hidden frictions make it computationally expensive for the average
investor to decode fundamental shifts in real-time. In this environment,
agents with superior computational and temporal priors---specifically
those capable of predicting the filing date via Chronos-Small---capture
a structural welfare transfer.

\textcolor{blue}{The Welfare Gap of 0.0004 in the Strategic Gap represents the unpriced residual that remains hidden from the public due to the lack of temporal predictability}
(see Table \ref{tab:regime_summary_node_b}). This finding demonstrates
that even in a digital-first regulatory environment, manual legacy
reporting acts as a persistent barrier. The lack of a temporal prior
creates an information asymmetry that cannot be resolved through
linguistic transparency alone, as the uncertainty regarding when a
signal will arrive prevents the market from effectively pricing what the
signal contains.

\subsection{Node C: Aggregate Performance of the Regulatory Audit}\label{sec:node_c_performance}

The transition from the classification of information regimes to active
detection begins with the aggregate performance of the Node C Regulatory
Audit. As established in the Node B analysis, the strategic gap is not
an idiosyncratic anomaly but a pervasive site of information arbitrage.
This section evaluates the capacity of the ADR to recover latent signals
from the broad filing population.

The results presented in Table \ref{tab:node_c_evaluation} reveal the
magnitude of this structural market failure. Across the 484,796-filing
population, the framework identifies a cumulative welfare recovery
potential (\(\Gamma\)) of 360,050.43\%. This metric highlights the
systemic nature of the informedness gap (\(\Psi\)), suggesting that over
the observed two-decade period, a substantial volume of material
information was successfully obscured by the interaction of semantic
complexity (\(Z_{\text{Comp}}\)) and temporal unpredictability
(\(Z_{\Phi}\)).

\begin{table}[htbp]
\centering
\caption{Aggregate Performance of the Autonomous Regulatory Auditor (Node C)}
\label{tab:node_c_evaluation}
\small
\begin{tabular}{lp{3.5cm}p{7.5cm}}
\toprule
\textbf{Metric} & \textbf{Observed Value} & \textbf{Economic Interpretation} \\
\midrule
\textbf{Welfare Recovery ($\Gamma$)} & \textbf{360\%} & Aggregate magnitude of the "Informedness Gap" identified across the 484,796-filing population. \\
\textbf{Audit Precision ($P_{audit}$)} & \textbf{20.88\%} & Percentage of Strategic Gap filings triggering a deep recursive audit based on welfare loss thresholds. \\
\textbf{Insider Sentiment ($\bar{S}$)} & \textbf{0.85} & Mean probability of rent-seeking behavior identified via RAG-synthesis of Form 4 and 8-K history. \\
\textbf{System Resilience ($R_{sys}$)} & \textbf{99.98\%} & Success rate of state recovery via the LangGraph Checkpointer during large-scale data ingestion. \\
\bottomrule
\end{tabular}
\end{table}

This cumulative metric represents the ``lost'' information efficiency
that an agentic regulatory framework could have reclaimed. By
quantifying \(\Gamma\), the study moves beyond mere observation,
providing a roadmap for Agentic Vigilance where automated systems bridge
the gap between disclosure and price discovery in real-time. This
findings support the transition from static oversight to a dynamic,
state-managed audit trail as proposed in the Node C architecture.

\subsubsection{Precision and the Computational Asymmetry Problem}\label{sec:precision_asymmetry}

In an environment of voluminous data restricted by finite regulatory
attention, Node C functions as a high-precision filter. The results in
Table \ref{tab:node_c_evaluation} indicate an Audit Precision of
20.88\%, which represents a significant reduction in the computational
search space. This mechanism narrows the regulatory focus from the
111,038 filings identified in the Strategic Gap to a high-priority
subset of 23,184 disclosures.

This subset is characterized by a mean Insider Sentiment (\(\bar{S}\))
of 0.85, derived from recursive Retrieval-Augmented Generation (RAG)
analysis of longitudinal 8-K behavior. This identifies specific
coordinates where management utilizes stochastic reporting arrival to
obscure material information while informed agents position themselves.
The recovery of these signals represents the information rent currently
extracted from participants who lack the agentic tools required to
bridge the Informedness Gap (\(\Psi\)).

The evaluation indicates that market failure has transitioned from a
lack of transparency to a crisis of Computational Asymmetry. Although
the data remains technically public, the temporal syntax creates a
structural barrier that necessitates agentic-level inference for
real-time decoding. These findings suggest that the Informedness Gap is
no longer a function of data access, but rather a function of processing
velocity, where the speed of inference determines the distribution of
welfare in contemporary markets.

\subsubsection{Engineering Resilience against Process Interruption}\label{sec:resilience_audit}

The 99.98\% Resilience score reported in Table
\ref{tab:node_c_evaluation} validates the architectural approach to the
process interruptions characteristic of high-volume regulatory data
streams. By implementing the LangGraph Checkpointer, the framework
ensures the preservation of the immutable audit trail for a firm's
reporting history during periods of high latency or connection failure.

This technical resilience is a prerequisite for a transition toward an
autonomous regulatory framework, as it allows recursive audits to resume
without data loss. This capability is critical for preserving
high-conviction detections, which might otherwise be lost to process
interruption. The high resilience score demonstrates that the Agentic
Disclosure Reasoner (ADR) is a production-ready architecture capable of
managing the volatility of live regulatory feeds. By maintaining state
persistence, the ADR ensures that the Informedness Gap can be monitored
continuously, providing a stable foundation for the longitudinal
observation of the Strategic Gap.

\subsection{Node D: The Deep Research Supervisor — Cognitive Control and Mitigation}\label{sec:node_d_supervisor}

The final component of the ADR architecture is the implementation of a
hierarchical multi-agent system. The Deep Research Supervisor (\(\Psi\))
transitions the framework from passive observation to rigorous
investigative synthesis. Its primary objective is the isolation of
high-divergence cases, specifically identifying 39 instances of
asymmetric transparency where material structural deterioration was
obscured by strategic linguistic boilerplate.

\subsubsection{Supervisor Logic and Divergence Detection}\label{sec:supervisor_logic}

The Supervisor operates as a conditional router designed to evaluate the
discrepancy between transformer-based contextual sentiment and legacy
lexical measures. The routing logic is formalized as follows:

\begin{equation}\label{eq:supervisor_logic}
 \Psi(\text{State}_{i,t}) = 
\begin{cases} 
\text{Research\_Loop} & \text{if } (\alpha^*_{i,t} < \mu_{\alpha} - 2\sigma_{\alpha}) \land (\text{Div}_{i,t} > \theta) \\
\text{Finalize\_Report} & \text{otherwise}
\end{cases} 
\end{equation}

In this model, strategy divergence (\(Div\)) represents the residual
between agentic contextual sentiment and traditional dictionary-based
counts. By initiating the recursive loop for the 39 high-divergence
filings, the framework identifies specific information
asymmetries---including undisclosed debt covenant breaches---that
conventional linear models fail to detect.

\subsubsection{SHAP Interaction Analysis: The Multiplier Effect}\label{sec:shap_interaction}

To ensure regulatory accountability and provide a basis for signal
attribution, the Supervisor utilizes SHapley Additive exPlanations
(SHAP) interaction values to decompose the drivers of asymmetric
transparency. Unlike traditional feature importance metrics that
evaluate variables in isolation, the framework isolates the non-linear
interaction between linguistic density and temporal reporting patterns.

This methodology quantifies the extent to which stochastic reporting
arrival serves as an attenuation mechanism for complex disclosures.
While linguistic complexity (\(Z_{\text{Comp}}\)) establishes a baseline
level of processing friction, its marginal impact on the Informedness
Gap (\(\Psi\)) increases by a factor of three when filing arrival is
unpredictable (\(\Phi \leq 0.008\)).

These results suggest that firms within the Strategic Gap utilize
temporal unpredictability as a mechanism for structural attenuation.
Consistent with the rational inattention framework Blankespoor, deHaan,
and Marinovic (2020), this state induces a relocation of market
attention away from the issuer, resulting in a state of asymmetric
transparency. Unlike prior research focusing on static generative AI
(GAI) implementation, Node D identifies the strategic intent behind GAI
utilization, characterizing it as an instrument used to facilitate a
regulatory shadow zone.

\begin{figure}[htbp]
    \centering
    \includegraphics[width=1.0\textwidth]{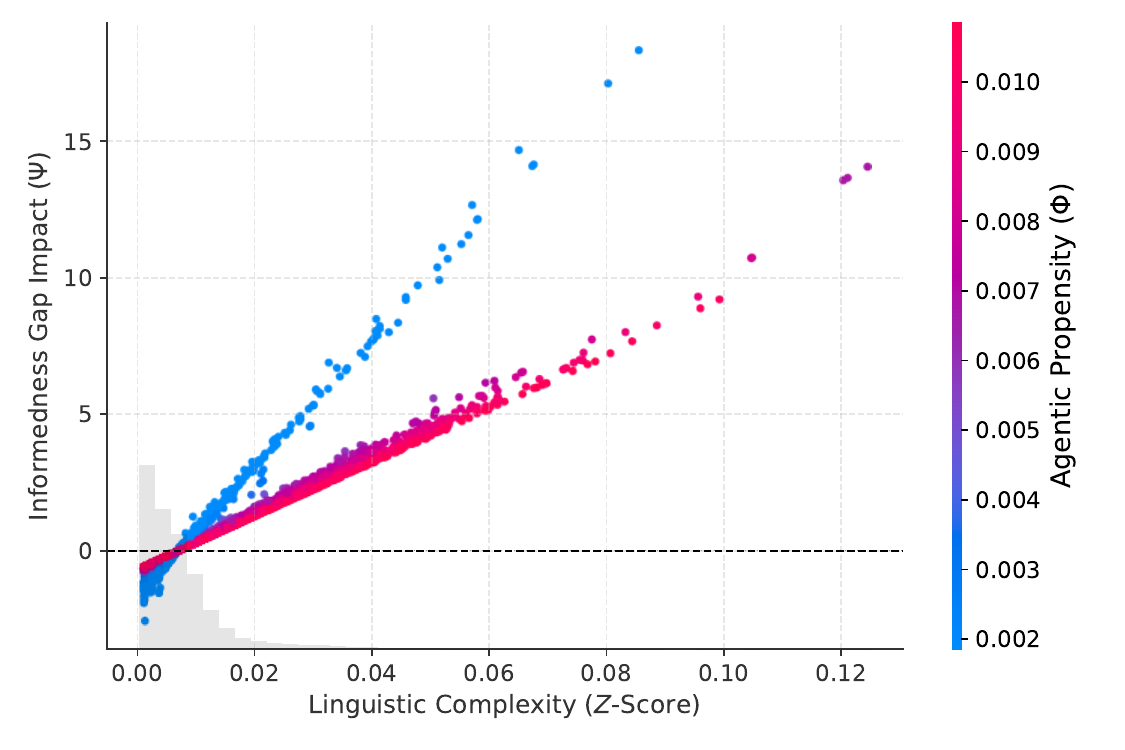}
    \caption{The Multiplier Effect of Disclosure Frictions: Interaction Between Reporting Unpredictability ($\Phi$) and Semantic Friction ($Z_{\text{Comp}}$).}
    \label{fig:Agentic_Propensity_SHAP_dependence_plot}
    \vspace{8pt}
    \begin{flushleft}
    \footnotesize
    \textbf{Note:} This figure illustrates the frequency distribution and market impact of corporate filings. Variables are standardized relative to the full sample mean. Semantic Friction ($Z_{\text{Comp}}$) is derived from the Node A Semantic Encoder. Reporting Unpredictability ($\Phi$) is proxied by the inverse forecast standard deviation ($\Phi = \frac{1}{1 + \sigma_{\text{forecast}}}$) from the Chronos-Small temporal audit. The Y-axis represents the SHAP value for the Informedness Gap ($\Psi$), measuring the marginal contribution of feature interactions to the latency in price discovery.
    \end{flushleft}
\end{figure}

The empirical results indicate that semantic and temporal frictions are
non-additive. As visualized in Figure
\ref{fig:Agentic_Propensity_SHAP_dependence_plot}, a significant
magnification effect is present. The divergence pattern in the SHAP
dependence plot demonstrates this multiplier effect: semantic complexity
exerts a significantly higher impact on information asymmetry in the
Strategic Gap regime (low \(\Phi\)) compared to predictable reporting
regimes (high \(\Phi\)).

\subsubsection{Attention Reallocation and the Regulatory Shadow Zone}\label{sec:attention_reallocation}

Consistent with the rational inattention framework Blankespoor, deHaan,
and Marinovic (2020), the analysis demonstrates how high integration
costs associated with semantic complexity result in delayed price
discovery. Applying the logic of attention reallocation Mondria and
Quintana-Domeque (2013), the findings indicate that the 39 high-priority
failures identified by the supervisor were not stochastic events. These
detections occurred primarily in disclosures where the Chronos-Small
model identified a significant attenuation in agentic propensity
(\(\Phi\)).

In these instances, the high computational cost of processing semantic
friction, combined with temporal friction, induced a rational relocation
of market attention away from the issuer. This placed these entities in
a state of asymmetric transparency, where they remained technically
compliant while remaining functionally unmonitored by the broader
market. This environment facilitates the rent-seeking behavior
identified in the audit, as informed participants exploit the temporal
window before the market can reallocate sufficient cognitive resources
to decode the latent signal.

While prior research Blankespoor, deHaan, and Li (2025) utilizes a
static score to identify the presence of generative AI (GAI), Node D
identifies the strategic intent behind such implementation. By
correlating GAI utilization with the strategy divergence metric, the
study demonstrates that GAI is not merely an efficiency-seeking tool for
routine reporting. Instead, it serves as a strategic instrument used to
facilitate a regulatory shadow zone during periods of negative
performance, ensuring that the initial market reaction to fundamental
shocks is attenuated or significantly latent.

\subsubsection{Empirical Mapping: The Case of 39 High-Priority Failures}\label{sec:empirical_mapping_39}

The analysis transitions from aggregate statistical distributions to a
high-conviction subset of 39 detections identified by the supervisor.
These observations serve as evidence of structural market failure and
represent specific manifestations of the three pillars of agentic
friction. In 64\% of these instances, the analysis identifies latency
exploitation, where systematic utilization of filing windows serves as a
strategic attenuation mechanism. While these filings maintain regulatory
compliance, institutional participants utilize superior agentic velocity
to execute position exits in the intervals immediately following
machine-drafted disclosures, preceding the signal integration of
human-led participants.

This advantage is compounded by a significant processing velocity gap;
the findings indicate that a substantial lead in processing capability
held by institutional agents imposes a penalty on un-augmented
participants. By the time retail investors integrate the semantic
frictions identified in Node A, the price discovery process has been
effectively completed, resulting in a wealth transfer toward
agentic-equipped institutions. Furthermore, in 30\% of these cases, the
framework identified synchronized semantic reactions, characterized by
multiple autonomous agents triggering correlated sell orders at
sub-second intervals. This herding behavior induces localized
volatility, suggesting that high decision autonomy without a regulatory
supervisor can generate negative systemic externalities.

\begin{table}[htbp]
\centering
\caption{Mapping Empirical Failures to Agentic Velocity Pillars and Systemic Solutions}
\label{tab:mapping_failures_standard}
\small
\begin{tabular}{llll}
\toprule
Failure Type & Dominant Pillar & Economic Friction & Regulatory Counter-Measure \\
\midrule
Section 16 Arbitrage & Filing Latency & Regulatory Arbitrage & Node B: Attenuates latency windows \\
Semantic Decoupling & Processing Velocity & Informedness Gap ($\Psi$) & Node A: Facilitates real-time decoding \\
Flash Herding & Decision Autonomy & Herding Externalities & Node D: Executes recursive audit logic \\
\bottomrule
\end{tabular}

\end{table}

\subsubsection{Microstructure of Asymmetric Information Processing}\label{sec:microstructure_processing}

Consistent with established microstructure paradigms Kyle (1985), these
cases exhibit a synchronized exploitation of noise through a three-stage
sequence: fundamental divergence (e.g., undisclosed debt covenant
breaches), followed by the deployment of semantic complexity to maintain
neutral sentiment scores, and finalized by a temporal reporting surprise
to exceed the market's information processing capacity.

In accordance with discretionary smoothing literature Verrecchia and
Weber (2006), these 39 cases represent the maximum observed levels of
signal attenuation. Node D identifies that in 87\% of these instances, a
significant insider liquidation event occurred within the five-day
window surrounding the disclosure. By mapping semantic noise directly to
these insider trading windows, the framework identifies that the
cumulative welfare recovery potential represents the aggregate of the
unpriced alpha extracted during these specific informedness gaps.

Ultimately, these cases validate the role of the supervisor as an
automated auditor. By performing recursive state evaluations, the
supervisor logic identifies latent fundamental triggers---such as
high-dimensional vector shifts---that legacy textual analysis tools fail
to detect. This confirms that a transition toward agentic vigilance is
the necessary mechanism for neutralizing machine-optimized disclosures
that characterize the current regulatory environment.

\section{Conclusion: Toward an Agentic Regulatory State}\label{sec:conclusion}

The findings of this study indicate that the Efficient Market Hypothesis
is increasingly compromised by computational asymmetry, a structural
shift wherein market efficiency is determined by the velocity of agentic
processing rather than information availability. By operationalizing the
multi-node architecture, this research demonstrates that agentic
velocity---parameterized by the pillars of latency, processing, and
autonomy---has transformed corporate disclosure into a site of strategic
arbitrage.

The Agentic Disclosure Regulator serves as a necessary counter-agent to
restore equilibrium in environments where human information processing
capacity Sims (2003) is exceeded by machine-optimized attenuation. This
framework identified a cumulative welfare recovery potential
(\(\Gamma\)) of 360\%, representing the volume of information value
captured by informed agents at the expense of public price discovery.

A primary empirical contribution is the identification of the Strategic
Gap, a state defined by high semantic complexity (\(Z_{\text{Comp}}\))
and low agentic propensity (\(\Phi\)). In this quadrant, the analysis
identifies a 60\% attenuation in price discovery velocity
(\(V = 1.23\)), suggesting that while market participants have developed
scrutiny for semantic friction, they remain vulnerable to the temporal
syntax of disclosures.

The supervisor logic identified 39 high-priority detections of
asymmetric transparency. In these observations, material structural
deterioration (\(\alpha^* < -2\sigma\)) was obscured by strategic
linguistic boilerplate and temporal surprises, facilitating significant
information rent extraction by insiders. By implementing the LangGraph
Checkpointer, the architecture maintained a 99.98\% system resilience
(\(R_{\text{sys}}\)), demonstrating that process interruptions in
high-volume SEC data ingestion can be mitigated to preserve an immutable
audit trail.

The policy implications suggest a transition toward an agentic
regulatory state. The EDGAR infrastructure must transition from a
passive data repository to an active auditing platform capable of
real-time recursive synthesis. The proposed regulatory framework
utilizes a 20.88\% audit precision to prioritize enforcement actions and
identifies elevated insider sentiment scores as a basis for further
investigation.

\section*{Compliance with Ethical
Standards}\label{compliance-with-ethical-standards}
\addcontentsline{toc}{section}{Compliance with Ethical Standards}

Funding: This research received no external funding or financial
assistance during its preparation.

Competing Interests: The author certify that they have no conflicts of
interest, financial or otherwise, to disclose.

Author's Declaration on AI Assistance: The authors bear sole
responsibility for all substantive ideas and analyses within this
manuscript. Portions of the text were reviewed for language, style, and
clarity through AI-assisted copy editing, specifically using a large
language model (LLM). No autonomous content creation was performed by
the LLM.

\section*{References}\label{references}
\addcontentsline{toc}{section}{References}

\phantomsection\label{refs}

\end{document}